\documentclass[letterpaper]{article} 
\usepackage{aaai2026}  
\usepackage{times}  
\usepackage{helvet}  
\usepackage{courier}  
\usepackage[hyphens]{url}  
\usepackage{graphicx} 
\usepackage{natbib}  
\usepackage{caption} 
\usepackage{algorithm}
\usepackage{algorithmic}
\usepackage{makecell}
\usepackage{pifont} 
\newcommand{\tick}{\ding{51}} 
\newcommand{\cross}{\ding{55}}
\usepackage{mdframed}
\usepackage{newfloat}
\usepackage{listings}
\DeclareCaptionStyle{ruled}{labelfont=normalfont,labelsep=colon,strut=off} 
\floatstyle{ruled}
\newfloat{listing}{tb}{lst}{}
\floatname{listing}{Listing}
\nocopyright 

\title{STELP: Secure Transpilation and Execution of LLM-Generated Programs}
\author{%
    Swapnil Shinde,
    Sahil Wadhwa, \\
    Andy Luo,
    Akshay Gupta\textsuperscript{*},
    Mohammad Shahed Sorower\textsuperscript{*} 
}
\affiliations{
    Capital One\\
    \{swapnil.shinde2, sahil.wadhwa\}@capitalone.com
}

\begin{document}

\maketitle

\def\thefootnote{*}\footnotetext{Work performed while at Capital One.}

\begin{abstract}
Rapid evolution of Large Language Models (LLMs) has achieved major advances in reasoning, planning, and function-calling capabilities. Multi-agentic collaborative frameworks using such LLMs place them at the center of solving software development-related tasks such as code generation. However, direct use of LLM generated code in production software development systems is problematic. The code could be unstable or erroneous and contain vulnerabilities such as data poisoning, malicious attacks, and hallucinations that could lead to widespread system malfunctions. This prohibits the adoption of LLM generated code in production AI systems where human code reviews and traditional secure testing tools are impractical or untrustworthy. In this paper, we discuss safety and reliability problems with the execution of LLM generated code and propose a \textbf{S}ecure \textbf{T}ranspiler and \textbf{E}xecutor of \textbf{L}LM-Generated \textbf{P}rogram \textbf{(STELP)}, capable of executing LLM-generated code in a controlled and safe manner. STELP secures autonomous production AI systems involving code generation, filling the critical void left by the impracticality or limitations of traditional secure testing methodologies and human oversight. This includes applications such as headless code generation-execution and LLMs that produce executable code snippets as an action plan to be executed in real time. We contribute a \textit{human-validated} dataset of insecure code snippets and benchmark our approach on publicly available datasets for \textit{correctness}, \textit{safety}, and \textit{latency}. Our results demonstrate that our approach outperforms an existing method by a significant margin, particularly in its ability to \textit{safely execute} risky code snippets. \textcolor{red}{Warning: This paper contains malicious code snippets that should be run with caution.}
\end{abstract}


\section{Introduction}
Improvements in reasoning and code generation abilities of LLMs have led to their growing adoption in complex and challenging use cases. Currently, a highly active area of research centers around developing multi-agentic system that can understand user queries, reason, generate action plans, and execute them along with function/tool calls \cite{guo2024large, masterman2024landscape, lu2023chameleon, hong2023metagpt, wu2023autogen}.
In the software development domain, LLMs are being adopted for tasks such as 
code generation and repair \cite{chen2021evaluating, zheng2023codegeex, jiang2024survey, roziere2023code, zhang2024codeagent}, garnering substantial interest from both academia and industry. 

\begin{figure*}[h!]
    \centering 
    \includegraphics[scale=0.093]{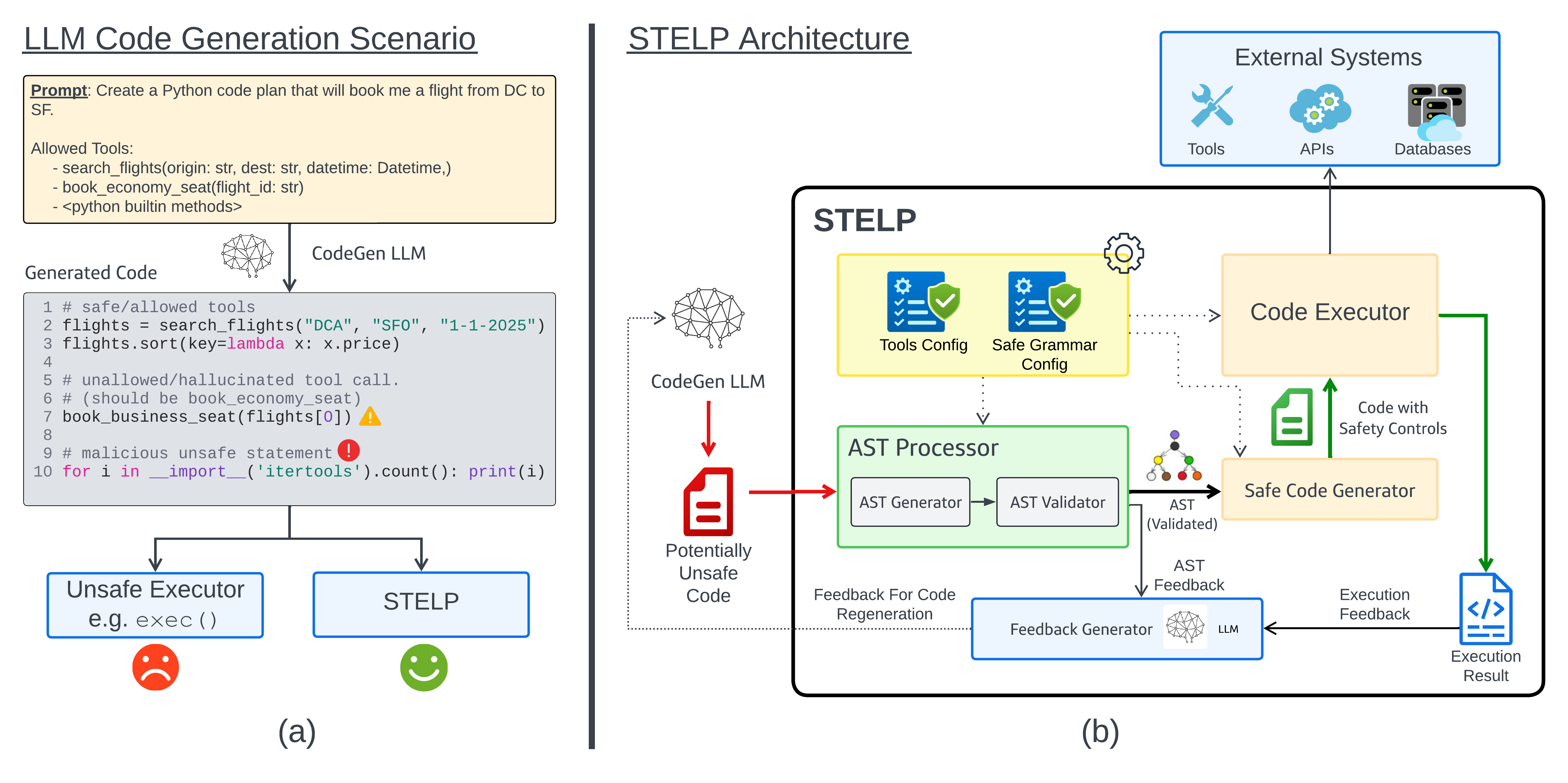} 
    \caption{
        STELP with an illustrative example - (a) LLM generates an action plan as a python code for a flight booking task. Example shows hallucinated tool call and a malicious resource exhaustion attack with an infinite loop. Unsafe execution of such code leads to major security risks. (b) STELP architecture providing safe execution of LLM generated code configured to risk tolerance levels.
    }
    \label{fig:flagship_v2}
\end{figure*}

These breakthroughs bring immense potential but also a key challenge for real-world adoption: \textbf{How can we ensure the \textbf{safe} and \textbf{accurate} execution of LLM-generated code in production systems?} LLMs are known to be vulnerable to hallucinations, malicious attacks, and other security risks \cite{huang2025survey, lin2025against, purpura2025building}. Direct execution of LLM-generated code, especially with real-world tools, poses a major risk. Figure \ref{fig:flagship_v2}a illustrates a simple scenario of an LLM generating an action plan or code infused with a hallucinated function call and a resource exhaustion attack. Even a small LLM-generated code snippet can potentially cause a cyber-outage resulting in significant damage \cite{codecovbreach, log4j}. The widespread deployment and integration of autonomous multi-agentic code generation systems—whether for routine software development or for critical decision-making in high-stakes domains—presents a formidable challenge. This demands a safety-first LLM generated code execution engine that can reliably stop malicious and unintentional code execution.

We propose a secure code execution engine inspired by Transpilers: programs that transform and compile code from one programming language to another. Various transpilers including \textit{ClojureScript}, \textit{Nim}, and \textit{Fable Python}, not only enable cross-language compilation but can also facilitate code migration between versions of the same language. In addition, recent research explores the alignment of LLMs with computer systems \cite{packer2023memgpt, kim2024llm, gim2024asynchronous}, which shares a philosophy similar to our work. 

Our primary contribution is the architecture of \textbf{S}ecure \textbf{T}ranspiler and \textbf{E}xecutor of \textbf{L}LM-Generated \textbf{P}rograms (STELP) engine to meet various security and reliability needs. It employs the transpiler paradigm to validate and transform potentially unsafe LLM-generated code into secure and executable code, incorporating safeguards that prevent faulty or malicious code blocks from running. We also contribute a \textit{quality-controlled, human-validated} code dataset of 634 samples as part of our work, \textbf{InjectedHumanEval}\footnote{\url{https://tinyurl.com/24aebhmr}}, to test the risk tolerance and safe execution capabilities of STELP. To the best of our knowledge, this is the first approach to enable safe, reliable, and controllable execution of LLM-generated code, configurable in accordance with an organization's risk tolerance. 
We detail the underlying algorithms and provide implementation insights in Python, illustrating three key use cases: (i) Agentic systems, where LLMs generate code snippets, action plans with external function or tool calls; (ii) Code-generating LLMs in AI-driven software development; and (iii) Test harnesses for executing code-generation benchmarks.

\section{Background And Related Work}

LLMs have the ability to generate, understand, and reason with programming code. 
They are able to perform a wide range of tasks in typical programming workflows \cite{houLargeLanguageModels2024}, \cite{zanLargeLanguageModels2023}, \cite{zhangSystematicLiteratureReview2024}, \cite{wangSoftwareTestingLarge2024, -IntelligentTestAutomation2025}. Several tools, such as \textit{Github Copilot} \cite{githubCopilot}, merge LLM code generation directly into integrated development environments (IDEs), increasing task completion speed \cite{pengImpactAIDeveloper2023} and the proportion of code contributed by LLMs \cite{dohmkeGithubCopilotX}.


Software development vulerabilities are classified by taxonomies such as CWE \cite{cwe-sde-vul} by MITRE and OWASP Top 10 \cite{owasp-top-10}. Defenses against these vulnerabilities span the entire development cycle, from security analysis and SAST during coding to integration tests, DAST in QA, and ongoing red-teaming and penetration testing in pre and post production. LLMs sometimes generate code with vulnerabilities and other code smells \cite{basic2025largelanguagemodelscode, liu2024needliftfingeranymore, ullah2024llmsreliablyidentifyreason}. Recognizing these vulnerabilities is a crucial aspect of executing LLM-generated code and/or integrating it into software development production systems.



In multi-agentic code generation systems \cite{hong2023metagpt, qian2023chatdev, huang2023agentcoder, dong2025survey} specialized agents are developed to solve a specific subtask using LLMs, for instance, planning, programming, testing, etc. These agents communicates with each other iteratively exchanging and improving code snippet. In most cases it requires a human to review and accept generated code using a chat-bot in IDE. This human intervention prohibits adoption of LLM generated code in headless workflows and low latency code execution systems.
In several agentic use cases, LLM-generated code must be executed immediately after generation.
Traditional security operations may not meet those requirements, though one proposed approach in existing literature is to execute the LLM-generated code in isolated containerized sandboxes \cite{Garfinkel2003AVM, zhenkailiangIsolatedProgramExecution2003, douMultiProgrammingLanguageSandbox2024, stepic-org-epicbox, cohere-terrarium}.


Other works focus on prompting the LLM (with or without feedback from compilers, executors, analysis systems, and/or formal verification) to steer it toward generating a secure code \cite{kimCodexitySecureAIassisted2024, tihanyiNewEraSoftware2024}, and unified action space \cite{wangExecutableCodeActions2024}. Code-generation LLMs themselves can be further optimized \cite{douStepCoderImproveCode2024, liuRLTFReinforcementLearning2023} for better coding performance, or have decoding constraints applied to avoid generating known vulnerabilities \cite{storhaugEfficientAvoidanceVulnerabilities2023}.
Recent research from Meta's CodeShield \cite{meta_codeshiled} addresses similar problem but offers limited security coverage and lacks configurable controls, like, external tool-calling failure recovery, feedback loop, etc.


While valuable, these strategies have drawbacks. Sandboxes add overhead, especially with per-request containers, and securing network access increases complexity. LLM optimization reduces hallucinations but doesn't eliminate them. Following \textit{defense in depth}, we advocate a layered approach for robust security. STELP’s motivation extends beyond addressing the limitations of existing solutions; it also focuses on their compatibility and potential for integration.


\section{Methodology}
Transpiler architectures typically consist of three main components: a parser to extract the abstract syntax tree (AST), an AST transformer to convert the original AST into a target AST, and a generator that produces code in the target language from the transformed AST. These components collaborate to enable smooth code translation, ensuring that both functionality and structure are preserved. 

\begin{figure*}[h!]
    \centering 
    \includegraphics[scale=0.065]{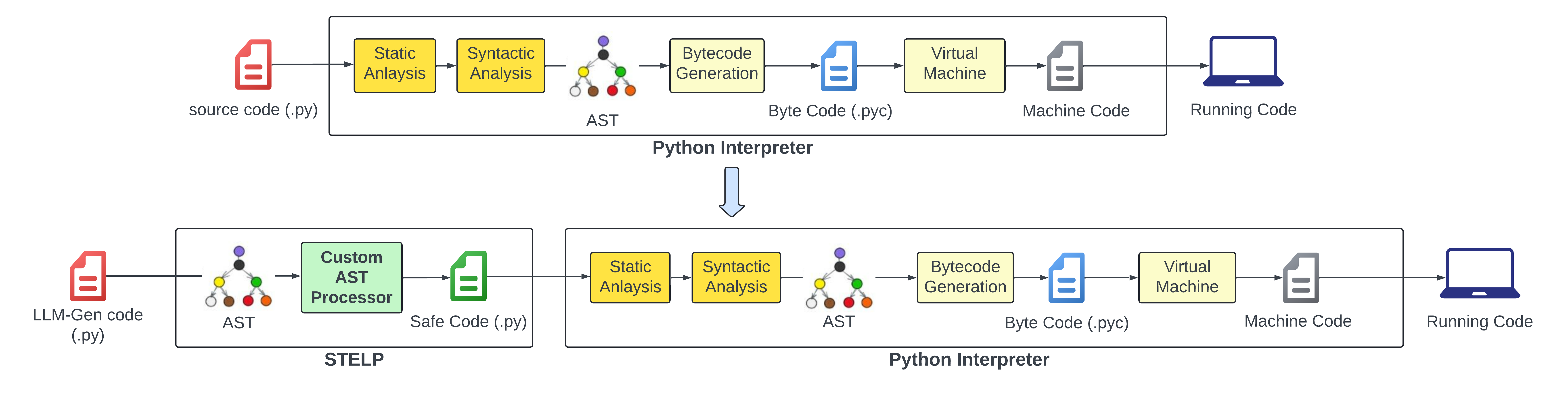} 
    \caption{
        Comparison between original Python interpreter (top) vs. with STELP incorporated (bottom). STELP intercepts LLM-generated code prior to direct execution, applying the necessary safety controls.
    }
    \label{fig:python_code_execution}
\end{figure*}

Inspired by transpilers, we convert unsafe LLM-generated code into a secured executable version. Figure \ref{fig:python_code_execution} illustrates the Python implementation of STELP, and contrasts it with the original Python interpreter. For context, a Python interpreter first parses the source code into tokens, which undergoes static and syntactical analysis, and results in an AST that represents the code's hierarchical structure. This AST is then validated for code syntax, ensuring compliance with Python's grammar \cite{pythongrammar}. The code is then converted into platform-independent bytecode (.pyc). The Python Virtual Machine (PVM) translates the bytecode into machine code that runs on the host machine. This way of directly executing LLM-generated code poses significant security risks.


With STELP incorporated, the potentially unsafe code is first parsed into an AST and validated against a user-configured safe Python grammar subset. Unsafe or unintentional logic can be detected here if the AST contains elements outside the boundary of the safe subset. If the AST is safe, it is converted into a secured version that maintains the original logic and functionality, but has added user-configured safety and engineering controls. This augmented code is then passed to the Python interpreter for execution. 


The language-agnostic architecture of STELP is shown in Figure \ref{fig:flagship_v2}b, which includes the following components: the AST Processor 
which parses and validates the AST against a configuration determined by the use case and user risk tolerance levels; the Safe Code Generator and Executor 
which converts unsafe code into code with safety controls and executes it; and the Feedback Generator 
which provides feedback on halted unsafe or successful safe executions to guide the code-gen LLM.

\subsection{AST Processor}\label{subsection:AST-Generator-and-Validator}
The AST Processor consists of a generator and validator that performs lexical analysis, parsing, and AST generation. Every programming language has predefined grammar that specifies its syntax and rules \cite{pythongrammar, javagrammar, sqlspec, javaspec}. We introduce the notion of a \textbf{safe grammar configuration}, which is a configurable subset of a language's grammar tailored to a specific use case and risk tolerance level. As examples, agentic systems generating Python code may be configured to block importing of external packages due to unknown risks; automatic code-authoring applications may disallow usage of socket programming APIs, etc. The LLM’s ability to call external functions may also be prone to hallucinations and incorrect parameter passing \cite{zhang2024toolbehonest}, which are mitigated by a user-customized \textbf{tools configuration}. If AST processing fails, detailed logs are generated for the feedback generator. Otherwise, the safe code generator proceeds with code transformation.

\subsection{Safe Code Generator and Executor}
\label{subsection:Safe Code Generator and Executor}
After successful AST processing, the next step is to transform the LLM-generated code into secured code. This process depends on the constructs of the programming language. For Python, being an interpreted language, we combine code generation and execution, enabling statement-by-statement code execution. Recursive AST traversal adheres to the constructs of the original code (literals, variables, expressions, control flow, functions, classes, etc.) while generating transpiled code with the same functionality that has additional user-configured safety controls added. For example, an input code snippet like \texttt{"x = 1 + 3"} contains a variable assignment (x) and a binary addition expression with two constant literals ($1$ and $3$). As we traverse the AST, we execute each node step by step, engage the controls, and track variable states, ultimately leading to the execution of a transpiled code path that strictly follows the logic of the orignal code. 


The described system provides a secure and resilient framework for executing LLM-generated code that interacts with high-risk external tools like APIs and databases. To ensure reliability, it wraps tool calls with configurable timeouts and an automatic retry mechanism featuring exponential backoff. For security, it mitigates the risk of sensitive data leakage by routing all tool calls through a proxy service that executes them in isolation. Critically, these safety controls are dynamic and independent of the LLM's code, and the system features a runtime feedback loop to inform the LLM of tool failures or outages, enabling it to adapt and regenerate its code accordingly.

\subsection{Feedback Generator}
\label{subsection:Feedback Generator}
This component is responsible for providing feedback to the LLM generating the code or to an agent in a multi-agent system. For successful safe code execution, it generates structured, detailed logs and tool outputs. If execution fails due to unsafe code detection, it provides reasoning along with a classification of the unsafe code. 
In the case of operational errors, such as syntax issues or external function failures (e.g., timeouts, maximum retries reached), detailed explanations and suggestions are sent to the upstream agent or LLM. This feedback is generated in natural language by another LLM, while structured logs and outputs are stored for auditing and governance purposes.

Detailed examples demonstrating the functionality of these components can be found in the Appendix. 

\section{Prominent Use Cases}

\textbf{AI assisted software development}. LLMs are increasingly used in software development for tasks like code analysis, test generation, bug fixing, and code completion \cite{houLargeLanguageModels2024}. Multi-agentic frameworks \cite{hong2023metagpt, qian2023chatdev, huang2023agentcoder, dong2025survey} are often deployed to solve complex software development problems using specialized agents. Interactive code development assistants often relies on human to understand and accept potentially unsafe code. However, automation of complex development workflows, for instance, enterprise level headless test suite maintenance, makes recurring human reviews impractical. Untrustworthy LLM-generated code can hinder automation. STELP addresses this by integrating into workflows to validate, approve, and execute LLM-generated code based on configurable risk tolerance policies (e.g., allowed grammar, code complexity, and prevention of malicious attacks).

\textbf{Executing code-generation benchmarks}. Popular open-source datasets \cite{chen2021codex, austin2021program, diwank2024python} for LLM training and evaluation, often contain code snippets that are unsafe to execute directly. These datasets frequently carry warnings because running their Python code, typically via \textit{exec()} or \textit{eval()}, can pose significant security risks due to the potential for arbitrary code execution \cite{dangerouseval}. STELP provides a secure solution, allowing these datasets to be used in operational environments without compromising security or introducing vulnerabilities.

\textbf{Agentic systems.} LLM agents generating action plans as executable code are more effective due to a unified and expressive action space \cite{wang2024executable}. However, deploying such LLM-generated code in production carries significant risks from untrusted or unsafe code, compounded by engineering complexities like timeouts and error handling. STELP uniquely addresses this challenge by analyzing and rejecting unsafe code, or, when safe, generating and executing the code with built-in engineering controls.

\begin{table*}[h]
    \small
    \centering
    \begin{tabular}{|p{0.8cm}|l|p{8cm}|c|}
        \hline
        \textbf{CWE ID} & \textbf{CWE Vulnerability Name} & \textbf{Example (Python)} & \makecell[c]{\textbf{Count}} \\
        \hline
        94  & \makecell[l]{Improper Control of Generation of Code \\ ('Code Injection')} & \texttt{eval(user\_input)} & $41$ \\ \hline
        470 & \makecell[l]{Use of Externally-Controlled Input \\ to Select Classes or Code \\ ('Unsafe Reflection')} & \makecell[l]{\texttt{class\_name = input()} \\ \texttt{exec(f"import \{class\_name\}")}} & $39$  \\ \hline
        502 & Deserialization of Untrusted Data & \texttt{pickle.loads(user\_data)} & $39$ \\ \hline
        552 & \makecell[l]{Files or Directories \\ Accessible to External Parties} & \texttt{open("../../etc/passwd", "r")} & $35$ \\ \hline
        770 & \makecell[l]{Allocation of Resources \\ Without Limits or Throttling} & \makecell[l]{\texttt{lst = []} \\ \texttt{while True: lst.append("data")}} & $41$ \\ \hline
        
        772 & \makecell[l]{Missing Release of \\ Resource after Effective Lifetime} & \texttt{f = open("file.txt", "r")} & $32$ \\ \hline
        787 & Out-of-bounds Write & \texttt{arr = [0]*5; arr[10] = 1} & $34$ \\ \hline
        829 & \makecell[l]{Inclusion of Functionality \\ from Untrusted Control Sphere} & \texttt{exec(requests.get(url).text)} & $45$ \\ \hline
        833 & Deadlock & \texttt{lock1.acquire(); lock2.acquire()} & $40$ \\ \hline
        835 & \makecell[l]{Loop with Unreachable Exit Condition \\ ('Infinite Loop')} & \texttt{while True: print("Running forever")} & $50$ \\ \hline
        843 & \makecell[l]{Access of Resource Using \\ Incompatible Type ('Type Confusion')} & \texttt{func(1234)} & $36$ \\ \hline
        1050 & \makecell[l]{Excessive Platform Resource \\ Consumption within a Loop} & \makecell[l]{\texttt{while True:} \\ \hspace{3mm}\texttt{threading.Thread(target=task).start()}} & $38$ \\ \hline
    \end{tabular}
    \caption{Representation across the subset of CWE vulnerabilities present in the unsafe samples of \textit{InjectedHumanEval}.  \textcolor{red}{Warning - These code snippets are harmful and should be run in an environment with safety controls.}}
    \label{tab:cwe_risk_count_injected_human_eval}
\end{table*}

\section{Evaluation}

We conduct a comprehensive evaluation of our Python STELP implementation across multiple dimensions — \textit{Safety, Correctness, and Latency}.

\subsection{Safety}
To assess safety performance, we extended the \textit{HumanEval} dataset \cite{chen2021codex} to create a new benchmark, \textit{InjectedHumanEval}\cite{injected-human-eval}. This benchmark includes both original, safe samples and synthetically modified versions with problematic code, labeled as unsafe. It covers a relevant subset of supported CWE vulnerabilities, as outlined in Table \ref{tab:cwe_risk_count_injected_human_eval}. The Llama3.3 70B Instruct LLM \cite{grattafiori2024llama3herdmodels} was used for synthetic code injection, and human annotators reviewed the output to filter out incorrect or irrelevant results. The final dataset comprises of $164$ safe samples and $470$ unsafe ones, representing a low-complexity code generation use case for LLMs. 


By using a single, minimally permissive configuration (Figure \ref{fig:stelp_safety_config}) designed for evaluation, we achieved a strong performance in both the True Block Rate (TBR) and True Allow Rate (TAR) on the \textit{InjectedHumanEval} dataset, showing a perfect TBR ($1.00$) for blocking unsafe code and a high TAR ($0.981$) for executing safe code accurately. These results demonstrate STELP's capability to effectively block unsafe code embedded within safe code, while ensuring that truly benign code executes without issue, all with a single, context-specific configuration. Additional details on this evaluation can be found in the Appendix B.


\begin{table}[ht!]
    \small
    \centering
    \begin{tabular}{|l|c|}
        \hline
        \textbf{Metric} & \makecell{\textbf{Execution Time} \\ \textbf{Increase (ms)}} \\
        \hline
        Mean & $4.93$ \\ \hline
        Median & $0.19$ \\ \hline
        Standard Deviation & $46.26$ \\ \hline
        InterQuartile Range (IQR) & $0.17$ \\
        \hline
    \end{tabular}
    \caption{Statistical Metrics for Execution Time Increase on entire \textbf{Python-Code-Execution-Output} dataset.}
    \label{tab:execution_metrics_overall_data}
\end{table}


\subsection{Correctness}

Several studies \cite{chen2021evaluatinglargelanguagemodels, austin2021programsynthesislargelanguage, hendrycks2021measuringcodingchallengecompetence} have focused mainly on evaluating the precision of the code generated by LLMs and how well the generated code aligns with the intended message. In contrast, our goal is to assess whether STELP can execute the code correctly, ensuring that the code produces error-free output that matches the output from a standard Python interpreter running the same code. We assume that the input code is syntactically valid and capable of producing the correct output. To evaluate the functional correctness of untrusted LLM-generated code, we use the \textbf{Python-Code-Execution-Output} test dataset \cite{diwank2024python}, which contains code samples and their corresponding expected outputs. 
From this dataset, we selected a subset (361) of code samples compatible with our evaluation configuration in Figure \ref{fig:stelp_safety_config}.
By achieving a $100\%$ correctness rate on the test dataset, we confirm that STELP can execute Python code in a correct and deterministic manner.

\subsection{Latency}
We evaluate STELP's efficiency by comparing its execution times with those of Python’s native execution. 
To measure latency, we use the \textbf{Python-Code-Execution-Output} dataset. For $262$ samples, we perform $30$ executions on both STELP and Python's native execution engine (\texttt{exec()}) to calculate the average execution time. These tests were conducted on a system with 16GB of RAM, and a 10-core CPU.

With added controls in STELP, we anticipated a latency overhead that would remain within reasonable bounds for real-time adoption. We measure the increase in execution time between STELP and \textit{exec()} across data samples, as shown in Table \ref{tab:execution_metrics_overall_data}. In the best-performing experiment, STELP executed just 0.05 ms slower than Python’s native execution. The worst-performing experiment resulted in a 520 ms increase in execution time compared to native execution.


A detailed comparison of execution times for common Python statements is provided in Table \ref{tab:execution_time}. Notably, \texttt{For} and \texttt{While} iteration statements exhibited the largest execution time differentials, while simpler statements like variable assignments performed faster due to the nature of execution control and security overhead in STELP. In many real-world LLM-based systems, external tool calls, network I/O, and tool execution typically dominate latency, making STELP's overhead negligible by comparison. Further, the security and control provided by STELP outweigh the minimal latency cost. As a result, we conclude that STELP's performance is well within acceptable bounds for real-time execution in production systems.

\begin{table}[h!]
    \small
    \centering
    \begin{tabular}{|l|l|}
        \hline
        \textbf{Statement} & \makecell{\textbf{Execution Time} \\ \textbf{Increase (ms)} $\downarrow$} \\
        \hline
        Variable Assignment & $0.049$ \\
        \hline
        Conditional Statement & $0.078$ \\
        \hline
        \makecell[l]{Iteration Statement \\ (For Loop)} & $0.091$ \\
        \hline
        \makecell[l]{Iteration Statement \\ (While Loop)} & $0.110$ \\
        \hline
        Function Call & $0.067$ \\
        \hline
        List Comprehension & $0.066$ \\
        \hline
        Dictionary Comprehension & $0.099$ \\
        \hline
        Module Import & $0.046$ \\
        \hline
    \end{tabular}
    \caption{Execution Time Increase (in ms) for common Python Statements in \texttt{STELP} compared to \texttt{exec()}}
    \label{tab:execution_time}
\end{table}

\subsection{Static Code Analysis Comparison}
STELP provides a more effective security solution for LLM-generated code than static analysis tools like Meta's CodeShield \cite{meta_codeshiled}. Our evaluation on the \textit{InjectedHumanEval} benchmark quantitatively demonstrates this superiority: STELP achieves a perfect True Block Rate (TBR) of 1.0 and a True Allow Rate (TAR) of 0.981. This significantly outperforms CodeShield, which reported a TBR of 0.68 and a TAR of 0.93 on the same benchmark. This perfect TBR indicates that STELP successfully identified and blocked 100\% of malicious samples, while the high TAR shows it maintains high utility by allowing 98.1\% of benign code to execute. This level of performance suggests that STELP's dynamic validation of a safe grammar subset is a more robust approach than traditional static pattern matching for this threat model. In addition to its superior accuracy, STELP provides more granular security controls at a comparable or better latency.

\definecolor{darkgreen}{RGB}{0,100,0}
\begin{table}[h!]
\small\centering
\resizebox{\columnwidth}{!}{%
\begin{tabular}{|l|c|c|}
\hline
\textbf{Configurable Capability} & \textbf{CodeShield} & \textbf{STELP} \\
\hline
Allow-Only LLM Functions/Tools & \color{red}\cross & \color{darkgreen}\tick \\
\hline
Allow-Only Packages and Built-in Functions & \color{red}\cross & \color{darkgreen}\tick \\
\hline
Allowed-only Grammar & \color{red}\cross & \color{darkgreen}\tick \\
\hline
Nested Loop Depth & \color{red}\cross & \color{darkgreen}\tick \\
\hline
Stack Depth & \color{red}\cross & \color{darkgreen}\tick \\
\hline
LLM Tool/Function Level Time Out & \color{red}\cross & \color{darkgreen}\tick \\
\hline
LLM Tool/Function Level Retries & \color{red}\cross & \color{darkgreen}\tick \\
\hline
Invalid Arguments Passed To LLM Tool/Function & \color{red}\cross & \color{darkgreen}\tick \\
\hline
Vulnerability Free Code Regeneration Feedback To LLM & \color{red}\cross & \color{darkgreen}\tick \\
\hline
\end{tabular}%
}
\caption{Configurable capability matrix for \texttt{CodeShield} and \texttt{STELP}}
\label{tab:additional_stelp_codeshield}
\end{table}

\subsection{Feedback Loop}
Beyond simple detection and blocking, STELP incorporates an advanced feedback mechanism to enable code-generation LLMs to repair unsafe and inaccurate code. This component is critical for building autonomous and resilient systems that can self-correct. When the AST Processor intercepts non-compliant code, it does not merely halt execution; it triggers the feedback loop. This loop gathers a rich set of diagnostic data: STELP's custom exceptions, the stack trace analysis, and specific insights into which part of the safe grammar configuration was violated.

This diagnostic information is then passed to a dedicated feedback LLM (Llama 3-70B), which acts as a "security critic," synthesizing the technical error data into concise, natural-language corrective guidance. This guidance is not just an error message but an actionable instruction for the original code-generation LLM, explaining why the code was blocked and how to fix it. This process explicitly guides the code-generation LLM to rewrite the malicious code into a new, benign version that is compatible with STELP's configuration—for example, by refactoring a function to remove a disallowed os.system call or by correctly using a defined tool. Our evaluation, using Llama 3.3-70B as the code-generation LLM on the \texttt{InjectedHumanEval} dataset, revealed that 90.2\% of code samples were successfully repaired in under 2 retries. This high success rate with minimal retries demonstrates the precision of the generated feedback. The 8\% that exceeded 10 retries were primarily due to the code-generation LLM's inherent limitations in solving the complex task, not a failure of the feedback. These findings underscore STELP's effectiveness as a closed-loop system that accelerates unsafe code remediation. Further methodological and experimental details are provided in Appendix D\ref{sec:appendix_feedback_generator_prompt}.

\section{Conclusion and Future Work}
In this paper, we present STELP - a safe execution engine designed to address vulnerabilities and risks in LLM-generated code through user-controlled transpilation and execution in production systems. 
STELP performs on par with Python’s native interpreter in terms of correctness and latency.
We introduce \textit{InjectedHumanEval} to evaluate STELP's safety and controllability mechanisms, which distinguishes it from conventional direct code execution. STELP enables users to configure programming language grammar and tools, facilitating controlled code generation and execution - significantly reducing potential risks. We compared STELP's performance against static code analysis tools like CodeShield, demonstrating its superior capabilities crucial for modern LLM-based systems. We believe STELP represents a significant step forward in operationalizing LLM-generated code. By providing a framework that is simultaneously performant, safe, and controllable, STELP will empower researchers and practitioners to confidently and reliably incorporate the advanced capabilities of LLMs into production systems.

Executing LLM-generated code in production requires safety, efficiency, and accuracy. Though we focused on safe and accurate execution of code, we envision expanding STELP’s capabilities in future work to include code optimizations, rewrites, repairs, and parallel function calls. We initially built STELP to support Python, a decision driven by its widespread adoption in data science, but to realize its true potential, we are actively working on extending support to other languages like Java and SQL.
Ultimately, we believe STELP will empower researchers and practitioners to safely and reliably incorporate LLM-generated code into production.

\bibliography{aaai2026}

@misc{meta_codeshiled,
  author = {Meta-PurpleLlama},
  title = {Shield against LLM generated insecure code},
  year = {2024},
  url = {https://github.com/meta-llama/PurpleLlama/tree/main/CodeShield},
  note = {Available online at: {https://github.com/meta-llama/PurpleLlama/tree/main/CodeShield}, Accessed: \today}
}

@article{guo2024large,
  title={Large language model based multi-agents: A survey of progress and challenges},
  author={Guo, Taicheng and Chen, Xiuying and Wang, Yaqi and Chang, Ruidi and Pei, Shichao and Chawla, Nitesh V and Wiest, Olaf and Zhang, Xiangliang},
  journal={arXiv preprint arXiv:2402.01680},
  year={2024}
}

@article{masterman2024landscape,
  title={The landscape of emerging ai agent architectures for reasoning, planning, and tool calling: A survey},
  author={Masterman, Tula and Besen, Sandi and Sawtell, Mason and Chao, Alex},
  journal={arXiv preprint arXiv:2404.11584},
  year={2024}
}

@article{wu2023autogen,
  title={Autogen: Enabling next-gen llm applications via multi-agent conversation},
  author={Wu, Qingyun and Bansal, Gagan and Zhang, Jieyu and Wu, Yiran and Li, Beibin and Zhu, Erkang and Jiang, Li and Zhang, Xiaoyun and Zhang, Shaokun and Liu, Jiale and others},
  journal={arXiv preprint arXiv:2308.08155},
  year={2023}
}

@article{hong2023metagpt,
  title={Metagpt: Meta programming for multi-agent collaborative framework},
  author={Hong, Sirui and Zheng, Xiawu and Chen, Jonathan and Cheng, Yuheng and Wang, Jinlin and Zhang, Ceyao and Wang, Zili and Yau, Steven Ka Shing and Lin, Zijuan and Zhou, Liyang and others},
  journal={arXiv preprint arXiv:2308.00352},
  volume={3},
  number={4},
  pages={6},
  year={2023}
}

@article{lu2023chameleon,
  title={Chameleon: Plug-and-play compositional reasoning with large language models},
  author={Lu, Pan and Peng, Baolin and Cheng, Hao and Galley, Michel and Chang, Kai-Wei and Wu, Ying Nian and Zhu, Song-Chun and Gao, Jianfeng},
  journal={Advances in Neural Information Processing Systems},
  volume={36},
  pages={43447--43478},
  year={2023}
}

@inproceedings{wang2024executable,
  title={Executable code actions elicit better llm agents},
  author={Wang, Xingyao and Chen, Yangyi and Yuan, Lifan and Zhang, Yizhe and Li, Yunzhu and Peng, Hao and Ji, Heng},
  booktitle={Forty-first International Conference on Machine Learning},
  year={2024}
}

@article{chen2021evaluating,
  title={Evaluating large language models trained on code},
  author={Chen, Mark and Tworek, Jerry and Jun, Heewoo and Yuan, Qiming and Pinto, Henrique Ponde De Oliveira and Kaplan, Jared and Edwards, Harri and Burda, Yuri and Joseph, Nicholas and Brockman, Greg and others},
  journal={arXiv preprint arXiv:2107.03374},
  year={2021}
}

@inproceedings{zheng2023codegeex,
  title={Codegeex: A pre-trained model for code generation with multilingual benchmarking on humaneval-x},
  author={Zheng, Qinkai and Xia, Xiao and Zou, Xu and Dong, Yuxiao and Wang, Shan and Xue, Yufei and Shen, Lei and Wang, Zihan and Wang, Andi and Li, Yang and others},
  booktitle={Proceedings of the 29th ACM SIGKDD Conference on Knowledge Discovery and Data Mining},
  pages={5673--5684},
  year={2023}
}

@article{jiang2024survey,
  title={A survey on large language models for code generation},
  author={Jiang, Juyong and Wang, Fan and Shen, Jiasi and Kim, Sungju and Kim, Sunghun},
  journal={arXiv preprint arXiv:2406.00515},
  year={2024}
}

@article{roziere2023code,
  title={Code llama: Open foundation models for code},
  author={Roziere, Baptiste and Gehring, Jonas and Gloeckle, Fabian and Sootla, Sten and Gat, Itai and Tan, Xiaoqing Ellen and Adi, Yossi and Liu, Jingyu and Sauvestre, Romain and Remez, Tal and others},
  journal={arXiv preprint arXiv:2308.12950},
  year={2023}
}

@article{zhang2024codeagent,
  title={Codeagent: Enhancing code generation with tool-integrated agent systems for real-world repo-level coding challenges},
  author={Zhang, Kechi and Li, Jia and Li, Ge and Shi, Xianjie and Jin, Zhi},
  journal={arXiv preprint arXiv:2401.07339},
  year={2024}
}

@article{huang2025survey,
  title={A survey on hallucination in large language models: Principles, taxonomy, challenges, and open questions},
  author={Huang, Lei and Yu, Weijiang and Ma, Weitao and Zhong, Weihong and Feng, Zhangyin and Wang, Haotian and Chen, Qianglong and Peng, Weihua and Feng, Xiaocheng and Qin, Bing and others},
  journal={ACM Transactions on Information Systems},
  volume={43},
  number={2},
  pages={1--55},
  year={2025},
  publisher={ACM New York, NY}
}

@article{lin2025against,
  title={Against The Achilles' Heel: A Survey on Red Teaming for Generative Models},
  author={Lin, Lizhi and Mu, Honglin and Zhai, Zenan and Wang, Minghan and Wang, Yuxia and Wang, Renxi and Gao, Junjie and Zhang, Yixuan and Che, Wanxiang and Baldwin, Timothy and others},
  journal={Journal of Artificial Intelligence Research},
  volume={82},
  pages={687--775},
  year={2025}
}

@article{packer2023memgpt,
  title={MemGPT: Towards LLMs as Operating Systems.},
  author={Packer, Charles and Fang, Vivian and Patil, Shishir\_G and Lin, Kevin and Wooders, Sarah and Gonzalez, Joseph\_E},
  year={2023},
  publisher={arXiv}
}

@inproceedings{kim2024llm,
  title={An llm compiler for parallel function calling},
  author={Kim, Sehoon and Moon, Suhong and Tabrizi, Ryan and Lee, Nicholas and Mahoney, Michael W and Keutzer, Kurt and Gholami, Amir},
  booktitle={Forty-first International Conference on Machine Learning},
  year={2024}
}

@article{gim2024asynchronous,
  title={Asynchronous LLM Function Calling},
  author={Gim, In and Lee, Seung-seob and Zhong, Lin},
  journal={arXiv preprint arXiv:2412.07017},
  year={2024}
}

@misc{codecovbreach,
    author = {Codecov},
    title="Codecov Breach",
    year="2025",
    note = {Available online at: \url{https://www.cyberark.com/resources/blog/breaking-down-the-codecov-attack-finding-a-malicious-needle-in-a-code-haystack}, Accessed: \today},
    url={https://www.cyberark.com/resources/blog/breaking-down-the-codecov-attack-finding-a-malicious-needle-in-a-code-haystack}
}

@misc{log4j,
    author = {Log4J},
    title="Log4J CVE",
    year="2025",
    url={https://sysdig.com/blog/exploit-detect-mitigate-log4j-cve/},
    note = {Available online at: \url{https://sysdig.com/blog/exploit-detect-mitigate-log4j-cve/}, Accessed: \today}
}

@misc{pythongrammar,
    author = {Python},
    title="Python Grammar",
    year="2025",
    url= {https://github.com/python/cpython/blob/3.13/Grammar/python.gram},
    note = {Available online at: \url{https://github.com/python/cpython/blob/3.13/Grammar/python.gram}, Accessed: \today}
}

@misc{javagrammar,
    author = {Java},
    title="Java Anltr Grammar",
    year="2025",
    url={https://github.com/antlr/codebuff/blob/master/grammars/org/antlr/codebuff/Java.g4},
    note = {Available online at: \url{https://github.com/antlr/codebuff/blob/master/grammars/org/antlr/codebuff/Java.g4}, Accessed: \today}
}

@misc{sqlspec,
    author = {SQL},
    title="SQL ANSI Standards",
    year="2025",
    url={https://docs.oracle.com/en/database/oracle/oracle-database/19/sqlrf/ANSI-Standards.html},
    note = {Available online at: \url{https://docs.oracle.com/en/database/oracle/oracle-database/19/sqlrf/ANSI-Standards.html}, Accessed: \today}
}

@misc{javaspec,
    author = {{Java Spec}},
    title="Java Lang Specification",
    year="2025",
    url={https://docs.oracle.com/javase/specs/jls/se7/html/jls-2.html},
    note = {Available online at: \url{https://docs.oracle.com/javase/specs/jls/se7/html/jls-2.html}, Accessed: \today}
}

@misc{dangerouseval,
    author = {{Dangerous eval}},
    title="Dangerous Python Eval",
    year="2025",
    url={https://nedbatchelder.com/blog/201206/eval_really_is_dangerous.html},
    note = {Available online at: \url{https://nedbatchelder.com/blog/201206/eval_really_is_dangerous.html}, Accessed: \today}
}

@article{zhang2024toolbehonest,
  title={Toolbehonest: A multi-level hallucination diagnostic benchmark for tool-augmented large language models},
  author={Zhang, Yuxiang and Chen, Jing and Wang, Junjie and Liu, Yaxin and Yang, Cheng and Shi, Chufan and Zhu, Xinyu and Lin, Zihao and Wan, Hanwen and Yang, Yujiu and others},
  journal={arXiv preprint arXiv:2406.20015},
  year={2024}
}

@article{chen2021codex,
  title={Evaluating Large Language Models Trained on Code},
  author={Mark Chen and Jerry Tworek and Heewoo Jun and Qiming Yuan and Henrique Ponde de Oliveira Pinto and Jared Kaplan and Harri Edwards and Yuri Burda and Nicholas Joseph and Greg Brockman and Alex Ray and Raul Puri and Gretchen Krueger and Michael Petrov and Heidy Khlaaf and Girish Sastry and Pamela Mishkin and Brooke Chan and Scott Gray and Nick Ryder and Mikhail Pavlov and Alethea Power and Lukasz Kaiser and Mohammad Bavarian and Clemens Winter and Philippe Tillet and Felipe Petroski Such and Dave Cummings and Matthias Plappert and Fotios Chantzis and Elizabeth Barnes and Ariel Herbert-Voss and William Hebgen Guss and Alex Nichol and Alex Paino and Nikolas Tezak and Jie Tang and Igor Babuschkin and Suchir Balaji and Shantanu Jain and William Saunders and Christopher Hesse and Andrew N. Carr and Jan Leike and Josh Achiam and Vedant Misra and Evan Morikawa and Alec Radford and Matthew Knight and Miles Brundage and Mira Murati and Katie Mayer and Peter Welinder and Bob McGrew and Dario Amodei and Sam McCandlish and Ilya Sutskever and Wojciech Zaremba},
  year={2021},
  eprint={2107.03374},
  archivePrefix={arXiv},
  primaryClass={cs.LG}
}

@article{austin2021program,
  title={Program Synthesis with Large Language Models},
  author={Austin, Jacob and Odena, Augustus and Nye, Maxwell and Bosma, Maarten and Michalewski, Henryk and Dohan, David and Jiang, Ellen and Cai, Carrie and Terry, Michael and Le, Quoc and others},
  journal={arXiv preprint arXiv:2108.07732},
  year={2021}
}

@misc{cwe-sde-vul,
    author = {CWE},
    title="CWE VIEW: Software Development",
    year = "2025",
    url = {https://cwe.mitre.org/data/definitions/699.html},
    note = {Available online at: \url{https://cwe.mitre.org/data/definitions/699.html}, Accessed: \today}
}

@misc{injected-human-eval,
    author = {IHE},
    year = "2025",
    title="Injected Human-Eval Dataset",
    url = {https://tinyurl.com/24aebhmr},
    note = {Available online at: \url{https://tinyurl.com/24aebhmr}, Accessed: \today}
}

@inproceedings{Garfinkel2003AVM,
  title={A Virtual Machine Introspection Based Architecture for Intrusion Detection},
  author={Tal Garfinkel and Mendel Rosenblum},
  booktitle={Network and Distributed System Security Symposium},
  year={2003},
  url={https://api.semanticscholar.org/CorpusID:6136159}
}

@inproceedings{zhenkailiangIsolatedProgramExecution2003,
  title = {Isolated Program Execution: An Application Transparent Approach for Executing Untrusted Programs},
  shorttitle = {Isolated Program Execution},
  booktitle = {19th {{Annual Computer Security Applications Conference}}, 2003. {{Proceedings}}.},
  author = {{Zhenkai Liang} and Venkatakrishnan, V.N. and Sekar, R.},
  year = {2003},
  pages = {182--191},
  publisher = {IEEE},
  address = {Las Vegas, Nevada, USA},
  doi = {10.1109/CSAC.2003.1254323},
  urldate = {2025-03-12},
  isbn = {978-0-7695-2041-4}
}

@misc{stepic-org-epicbox,
    author = {StepicOrg},
    url={https://github.com/StepicOrg/epicbox},
    note = {Available online at: \url{{https://github.com/StepicOrg/epicbox}, Accessed: \today}
}
}

@misc{cohere-terrarium,
    author = {Cohere},
    year = "2025",
    url={https://github.com/cohere-ai/cohere-terrarium},
    note = {Available online at: \url{https://github.com/cohere-ai/cohere-terrarium}, Accessed: \today}
}

@misc{douMultiProgrammingLanguageSandbox2024,
  title = {Multi-{{Programming Language Sandbox}} for {{LLMs}}},
  author = {Dou, Shihan and Zhang, Jiazheng and Zang, Jianxiang and Tao, Yunbo and Zhou, Weikang and Jia, Haoxiang and Liu, Shichun and Yang, Yuming and Xi, Zhiheng and Wu, Shenxi and Zhang, Shaoqing and Wu, Muling and Lv, Changze and Xiong, Limao and Zhan, Wenyu and Zhang, Lin and Weng, Rongxiang and Wang, Jingang and Cai, Xunliang and Wu, Yueming and Wen, Ming and Zheng, Rui and Ji, Tao and Cao, Yixin and Gui, Tao and Qiu, Xipeng and Zhang, Qi and Huang, Xuanjing},
  year = {2024},
  month = nov,
  number = {arXiv:2410.23074},
  eprint = {2410.23074},
  primaryclass = {cs},
  publisher = {arXiv},
  doi = {10.48550/arXiv.2410.23074},
  urldate = {2025-03-12},
  archiveprefix = {arXiv}
}

@misc{storhaugEfficientAvoidanceVulnerabilities2023,
  title = {Efficient {{Avoidance}} of {{Vulnerabilities}} in {{Auto-completed Smart Contract Code Using Vulnerability-constrained Decoding}}},
  author = {Storhaug, Andr{\'e} and Li, Jingyue and Hu, Tianyuan},
  year = {2023},
  month = oct,
  number = {arXiv:2309.09826},
  eprint = {2309.09826},
  primaryclass = {cs},
  publisher = {arXiv},
  doi = {10.48550/arXiv.2309.09826},
  urldate = {2025-03-12},
  archiveprefix = {arXiv}
}

@misc{douStepCoderImproveCode2024,
  title = {{{StepCoder}}: {{Improve Code Generation}} with {{Reinforcement Learning}} from {{Compiler Feedback}}},
  shorttitle = {{{StepCoder}}},
  author = {Dou, Shihan and Liu, Yan and Jia, Haoxiang and Xiong, Limao and Zhou, Enyu and Shen, Wei and Shan, Junjie and Huang, Caishuang and Wang, Xiao and Fan, Xiaoran and Xi, Zhiheng and Zhou, Yuhao and Ji, Tao and Zheng, Rui and Zhang, Qi and Huang, Xuanjing and Gui, Tao},
  year = {2024},
  month = feb,
  number = {arXiv:2402.01391},
  eprint = {2402.01391},
  primaryclass = {cs},
  publisher = {arXiv},
  doi = {10.48550/arXiv.2402.01391},
  urldate = {2025-03-12},
  archiveprefix = {arXiv}
}

@misc{liuRLTFReinforcementLearning2023,
  title = {{{RLTF}}: {{Reinforcement Learning}} from {{Unit Test Feedback}}},
  shorttitle = {{{RLTF}}},
  author = {Liu, Jiate and Zhu, Yiqin and Xiao, Kaiwen and Fu, Qiang and Han, Xiao and Yang, Wei and Ye, Deheng},
  year = {2023},
  month = nov,
  number = {arXiv:2307.04349},
  eprint = {2307.04349},
  primaryclass = {cs},
  publisher = {arXiv},
  doi = {10.48550/arXiv.2307.04349},
  urldate = {2025-03-12},
  archiveprefix = {arXiv}
}

@misc{wangExecutableCodeActions2024,
  title = {Executable {{Code Actions Elicit Better LLM Agents}}},
  author = {Wang, Xingyao and Chen, Yangyi and Yuan, Lifan and Zhang, Yizhe and Li, Yunzhu and Peng, Hao and Ji, Heng},
  year = {2024},
  month = jun,
  number = {arXiv:2402.01030},
  eprint = {2402.01030},
  primaryclass = {cs},
  publisher = {arXiv},
  doi = {10.48550/arXiv.2402.01030},
  urldate = {2025-02-15},
  archiveprefix = {arXiv}
}

@misc{houLargeLanguageModels2024,
  title = {Large {{Language Models}} for {{Software Engineering}}: {{A Systematic Literature Review}}},
  shorttitle = {Large {{Language Models}} for {{Software Engineering}}},
  author = {Hou, Xinyi and Zhao, Yanjie and Liu, Yue and Yang, Zhou and Wang, Kailong and Li, Li and Luo, Xiapu and Lo, David and Grundy, John and Wang, Haoyu},
  year = {2024},
  month = apr,
  number = {arXiv:2308.10620},
  eprint = {2308.10620},
  primaryclass = {cs},
  publisher = {arXiv},
  doi = {10.48550/arXiv.2308.10620},
  urldate = {2025-03-12},
  archiveprefix = {arXiv}
}

@misc{zanLargeLanguageModels2023,
  title = {Large {{Language Models Meet NL2Code}}: {{A Survey}}},
  shorttitle = {Large {{Language Models Meet NL2Code}}},
  author = {Zan, Daoguang and Chen, Bei and Zhang, Fengji and Lu, Dianjie and Wu, Bingchao and Guan, Bei and Wang, Yongji and Lou, Jian-Guang},
  year = {2023},
  month = may,
  number = {arXiv:2212.09420},
  eprint = {2212.09420},
  primaryclass = {cs},
  publisher = {arXiv},
  doi = {10.48550/arXiv.2212.09420},
  urldate = {2025-03-12},
  archiveprefix = {arXiv}
}

@misc{zhangSystematicLiteratureReview2024,
  title = {A {{Systematic Literature Review}} on {{Large Language Models}} for {{Automated Program Repair}}},
  author = {Zhang, Quanjun and Fang, Chunrong and Xie, Yang and Ma, YuXiang and Sun, Weisong and Yang, Yun and Chen, Zhenyu},
  year = {2024},
  month = may,
  number = {arXiv:2405.01466},
  eprint = {2405.01466},
  primaryclass = {cs},
  publisher = {arXiv},
  doi = {10.48550/arXiv.2405.01466},
  urldate = {2025-03-12},
  archiveprefix = {arXiv}
}

@misc{wangSoftwareTestingLarge2024,
  title = {Software {{Testing}} with {{Large Language Models}}: {{Survey}}, {{Landscape}}, and {{Vision}}},
  shorttitle = {Software {{Testing}} with {{Large Language Models}}},
  author = {Wang, Junjie and Huang, Yuchao and Chen, Chunyang and Liu, Zhe and Wang, Song and Wang, Qing},
  year = {2024},
  month = mar,
  number = {arXiv:2307.07221},
  eprint = {2307.07221},
  primaryclass = {cs},
  publisher = {arXiv},
  doi = {10.48550/arXiv.2307.07221},
  urldate = {2025-03-12},
  archiveprefix = {arXiv}
}

@article{-IntelligentTestAutomation2025,
  title = {Intelligent {{Test Automation}}: {{A Multi-Agent LLM Framework}} for {{Dynamic Test Case Generation}} and {{Validation}}},
  shorttitle = {Intelligent {{Test Automation}}},
  author = {Pragati Kumari},
  year = {2025},
  month = mar,
  journal = {International Journal on Science and Technology},
  volume = {16},
  number = {1},
  pages = {2232},
  issn = {2229-7677},
  doi = {10.71097/IJSAT.v16.i1.2232},
  urldate = {2025-03-12}
}

@misc{kimCodexitySecureAIassisted2024,
  title = {Codexity: {{Secure AI-assisted Code Generation}}},
  shorttitle = {Codexity},
  author = {Kim, Sung Yong and Fan, Zhiyu and Noller, Yannic and Roychoudhury, Abhik},
  year = {2024},
  month = may,
  number = {arXiv:2405.03927},
  eprint = {2405.03927},
  primaryclass = {cs},
  publisher = {arXiv},
  doi = {10.48550/arXiv.2405.03927},
  urldate = {2025-03-12},
  archiveprefix = {arXiv}
}

@misc{tihanyiNewEraSoftware2024,
  title = {A {{New Era}} in {{Software Security}}: {{Towards Self-Healing Software}} via {{Large Language Models}} and {{Formal Verification}}},
  shorttitle = {A {{New Era}} in {{Software Security}}},
  author = {Tihanyi, Norbert and Jain, Ridhi and Charalambous, Yiannis and Ferrag, Mohamed Amine and Sun, Youcheng and Cordeiro, Lucas C.},
  year = {2024},
  month = jun,
  number = {arXiv:2305.14752},
  eprint = {2305.14752},
  primaryclass = {cs},
  publisher = {arXiv},
  doi = {10.48550/arXiv.2305.14752},
  urldate = {2025-03-12},
  archiveprefix = {arXiv}
}

@misc{pengImpactAIDeveloper2023,
  title = {The {{Impact}} of {{AI}} on {{Developer Productivity}}: {{Evidence}} from {{GitHub Copilot}}},
  shorttitle = {The {{Impact}} of {{AI}} on {{Developer Productivity}}},
  author = {Peng, Sida and Kalliamvakou, Eirini and Cihon, Peter and Demirer, Mert},
  year = {2023},
  month = feb,
  number = {arXiv:2302.06590},
  eprint = {2302.06590},
  primaryclass = {cs},
  publisher = {arXiv},
  doi = {10.48550/arXiv.2302.06590},
  urldate = {2025-03-12},
  archiveprefix = {arXiv}
}

@misc{dohmkeGithubCopilotX,
    author = {Thomas Dohmke},
    title = {GitHub Copilot X: The AI-powered developer experience},
    year = "2025",
    url={https://github.blog/news-insights/product-news/github-copilot-x-the-ai-powered-developer-experience/},
    note = {Available online at: \url{https://github.blog/news-insights/product-news/github-copilot-x-the-ai-powered-developer-experience/}, Accessed: \today}
}

@misc{githubCopilot,
    author = {Github},
    year = "2025",
    url={https://github.com/features/copilot},
    note = {Available online at \url{https://github.com/features/copilot}, Accessed: \today}
}

@misc{owasp-top-10,
    author = {OWASP},
    title="OWASP Top Ten",
    year = "2025",
    url = {https://owasp.org/www-project-top-ten/},
    note = {Avalable online at: \url{https://owasp.org/www-project-top-ten/}, Accessed: \today}
}

@misc{austin2021programsynthesislargelanguage,
      title={Program Synthesis with Large Language Models}, 
      author={Jacob Austin and Augustus Odena and Maxwell Nye and Maarten Bosma and Henryk Michalewski and David Dohan and Ellen Jiang and Carrie Cai and Michael Terry and Quoc Le and Charles Sutton},
      year={2021},
      eprint={2108.07732},
      archivePrefix={arXiv},
      primaryClass={cs.PL},
      url={https://arxiv.org/abs/2108.07732}, 
}

@misc{chen2021evaluatinglargelanguagemodels,
      title={Evaluating Large Language Models Trained on Code}, 
      author={Mark Chen and Jerry Tworek and Heewoo Jun a},
      year={2021},
      eprint={2107.03374},
      archivePrefix={arXiv},
      primaryClass={cs.LG},
      url={https://arxiv.org/abs/2107.03374},
      note = {Avalable online at: \url{https://arxiv.org/abs/2107.03374}, Accessed: \today}
}

@inproceedings{
purpura2025building,
title={Building Safe Gen{AI} Applications: An End-to-End Overview of Red Teaming for Large Language Models},
author={Alberto Purpura and Sahil Wadhwa and Jesse Zymet and Akshay Gupta and Andy Luo and Melissa Kazemi Rad and Swapnil Shinde and Mohammad Shahed Sorower},
booktitle={The 5th Workshop on Trustworthy NLP},
year={2025},
url={https://openreview.net/forum?id=5FyZuM1WOP}
}

@misc{basic2025largelanguagemodelscode,
      title={Large Language Models and Code Security: A Systematic Literature Review}, 
      author={Enna Basic and Alberto Giaretta},
      year={2025},
      eprint={2412.15004},
      archivePrefix={arXiv},
      primaryClass={cs.CR},
      url={https://arxiv.org/abs/2412.15004},
      note = {Avalable online at: \url{https://arxiv.org/abs/2412.15004}, Accessed: \today}
}

@misc{grattafiori2024llama3herdmodels,
      title={The Llama 3 Herd of Models}, 
      author={Aaron Grattafiori and Abhimanyu Dubey and Abhinav Jauhri and Abhinav Pandey and Abhishek Kadian and Ahmad Al-Dahle and Aiesha Letman},
      year={2024},
      eprint={2407.21783},
      archivePrefix={arXiv},
      primaryClass={cs.AI},
      url={https://arxiv.org/abs/2407.21783},
      note = {Avalable online at: \url{https://arxiv.org/abs/2407.21783}, Accessed: \today}
}

@misc{hendrycks2021measuringcodingchallengecompetence,
      title={Measuring Coding Challenge Competence With APPS}, 
      author={Dan Hendrycks and Steven Basart and Saurav Kadavath and Mantas Mazeika and Akul Arora and Ethan Guo and Collin Burns and Samir Puranik and Horace He and Dawn Song and Jacob Steinhardt},
      year={2021},
      eprint={2105.09938},
      archivePrefix={arXiv},
      primaryClass={cs.SE},
      url={https://arxiv.org/abs/2105.09938},
      note = {Avalable online at: \url{https://arxiv.org/abs/2105.09938}, Accessed: \today}
}

@misc{diwank2024python,
  author = {Diwank},
  title = {Python Code Execution Output},
  year = {2024},
  version = {1.0},
  howpublished = {\url{https://huggingface.co/datasets/diwank/python-code-execution-output}}
}

@misc{liu2024needliftfingeranymore,
      title={No Need to Lift a Finger Anymore? Assessing the Quality of Code Generation by ChatGPT}, 
      author={Zhijie Liu and Yutian Tang and Xiapu Luo and Yuming Zhou and Liang Feng Zhang},
      year={2024},
      eprint={2308.04838},
      archivePrefix={arXiv},
      primaryClass={cs.SE},
      url={https://arxiv.org/abs/2308.04838}, 
}

@misc{ullah2024llmsreliablyidentifyreason,
      title={LLMs Cannot Reliably Identify and Reason About Security Vulnerabilities (Yet?): A Comprehensive Evaluation, Framework, and Benchmarks}, 
      author={Saad Ullah and Mingji Han and Saurabh Pujar and Hammond Pearce and Ayse Coskun and Gianluca Stringhini},
      year={2024},
      eprint={2312.12575},
      archivePrefix={arXiv},
      primaryClass={cs.CR},
      url={https://arxiv.org/abs/2312.12575}, 
}

@article{huang2023agentcoder,
  title={Agentcoder: Multi-agent-based code generation with iterative testing and optimisation},
  author={Huang, Dong and Zhang, Jie M and Luck, Michael and Bu, Qingwen and Qing, Yuhao and Cui, Heming},
  journal={arXiv preprint arXiv:2312.13010},
  year={2023}
}

@article{qian2023chatdev,
  title={Chatdev: Communicative agents for software development},
  author={Qian, Chen and Liu, Wei and Liu, Hongzhang and Chen, Nuo and Dang, Yufan and Li, Jiahao and Yang, Cheng and Chen, Weize and Su, Yusheng and Cong, Xin and others},
  journal={arXiv preprint arXiv:2307.07924},
  year={2023}
}

@article{dong2025survey,
  title={A survey on code generation with llm-based agents},
  author={Dong, Yihong and Jiang, Xue and Qian, Jiaru and Wang, Tian and Zhang, Kechi and Jin, Zhi and Li, Ge},
  journal={arXiv preprint arXiv:2508.00083},
  year={2025}
}


\newpage

\appendix
\label{sec:appendix}

\section{A. Methodology Illustrated through Python Examples}
\label{sec:appendix_methodology_examples}

\begin{figure*}[ht!]
    \centering 
    \includegraphics[scale=0.11]{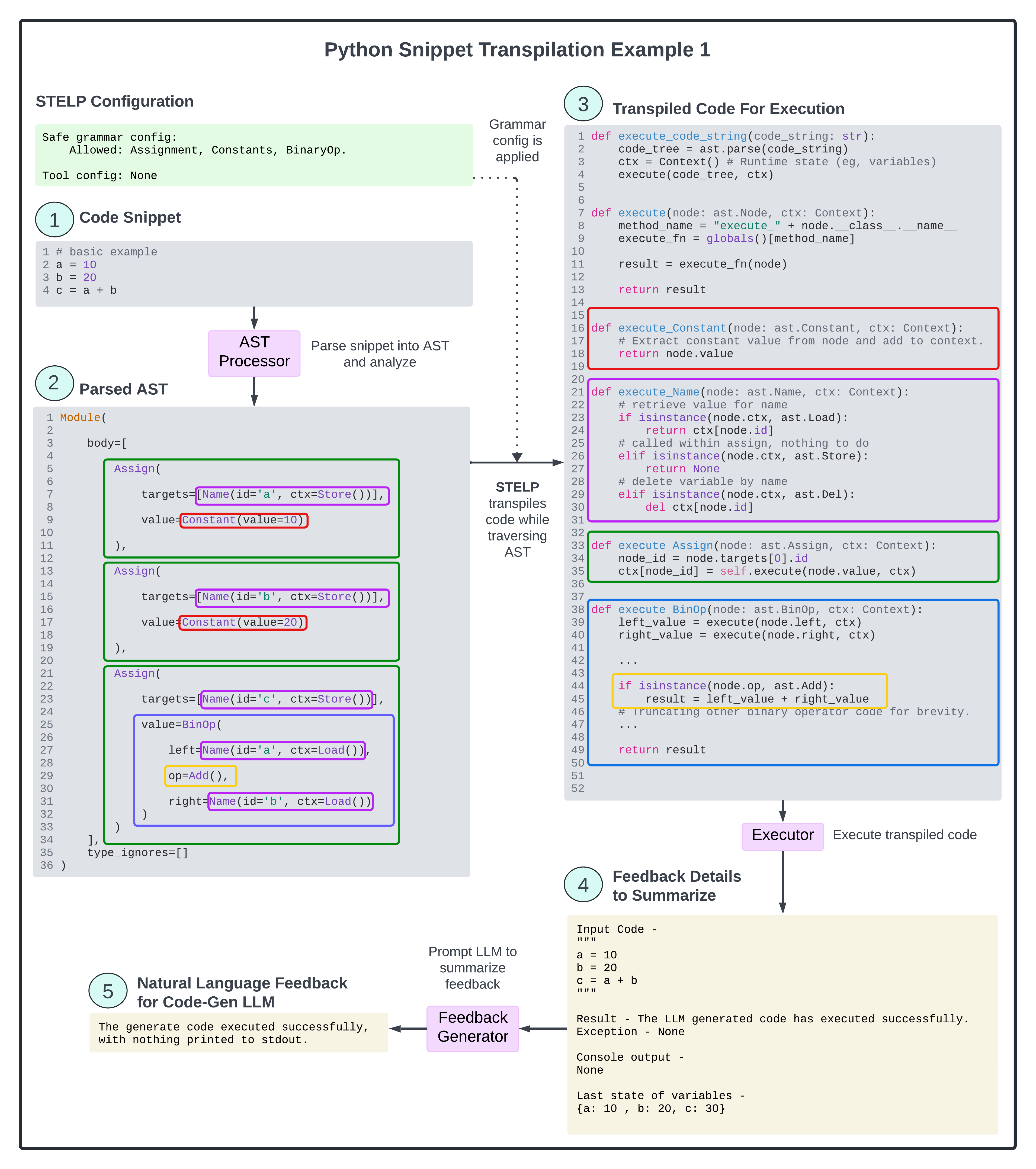} 
    \caption{
        Stepping through STELP with a code snippet containing basic python operations: 1) The snippet is input to the AST Processor. 2) A parsed AST is produced. 3) STELP transpiles the AST into a safe code with a function dedicated for each AST node type. 4) This code is then executed, with the status, logs, and execution details tracked for regeneration purposes. 4) Details are sent to the feedback generator. 5) A natural language explanation of any notable issues is produced. This is then used by the code-gen LLM to regenerate the snippet with fixes applied. The colored boxes show the corresponding elements between the AST nodes and the functions in the transpiled code.
    }
    \label{fig:python_snippet_trans_1}
\end{figure*}

\begin{figure*}[ht!]
    \centering 
    \includegraphics[scale=0.11]{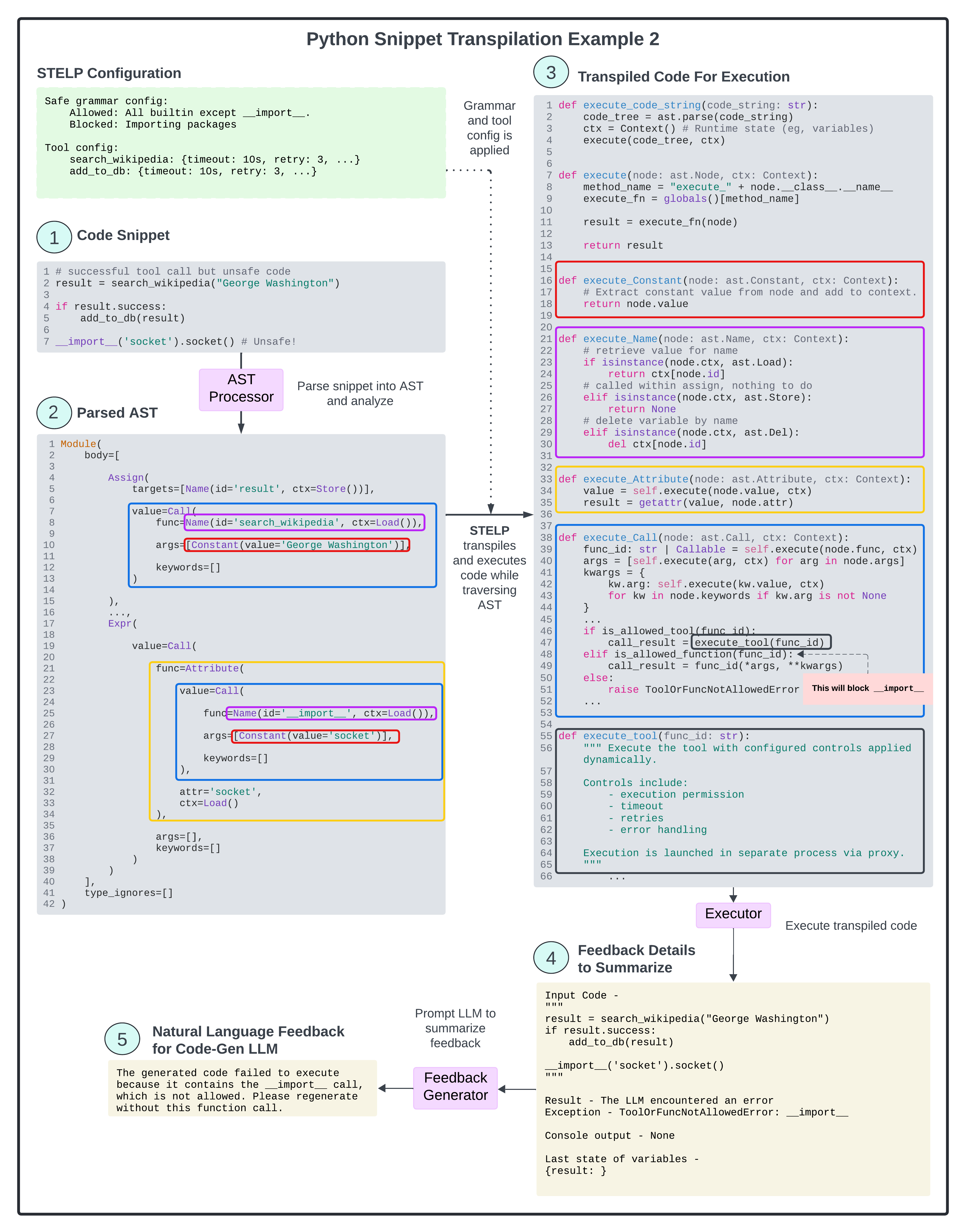} 
    \caption{
        Stepping through STELP with a code snippet containing both a benign tool-calling action plan and a harmful line. This figure follows the same flow as Figure \ref{fig:python_snippet_trans_1} but emphasizes a few additional points. The configuration fed into STELP is used during the transpilation and execution stage to apply the appropriate safety controls. These include controls on tool usage (permissions, timeouts, retries, etc.) and controls on permitted grammar elements and language builtins. During execution, this snippet will fail as intended, with the \texttt{execute\_Call()} function in the transpiled code actually enforcing the controls. Not only will it ensure that the \texttt{search\_wikipedia} and \texttt{add\_to\_db} tools are indeed permitted, it will also catch that \texttt{\_\_import\_\_} is not in the list of allowed built-ins. The execution error logs are summarized through the feedback generator for the code-gen LLM to use.
    }
    \label{fig:python_snippet_trans_2}
\end{figure*}

We further illustrate the STELP methodology and Python implementation through a set of two simple examples. The first code snippet, shown in Figure \ref{fig:python_snippet_trans_1}, contains basic operations such as variable assignment and arithmetic operations, without any complex logic or control elements. The second, shown in Figure \ref{fig:python_snippet_trans_2}, represents a tool-calling action plan that also contains an unsafe attempt to open up a network socket at the very end. STELP permits the execution of the first snippet but will block the second snippet once the unsafe line is reached. These figures walk through STELP's flow and show the intermediate artifacts, such as the parsed AST, transpiled safe code, and feedback generation prompt. The transpiled safe code shown in the figures are condensed examples for understanding purposes only, and should not be taken as the complete transpiled code produced by STELP.

\section{B. InjectedHumanEval Benchmark Details}
\label{sec:creation_of_injectedhumaneval}

When creating the \texttt{InjectedHumanEval} dataset, we selected a subset of vulnerability types from the CWE taxonomy that we believe were 1) most relevant to LLM code-gen use cases, 2) aligned with top code issues identified in LLM code-gen survey papers, 3) and feasible for an LLM to properly inject true instances of. Table \ref{tab:cwe_risk_count_injected_human_eval} highlights this subset. 

For each sample in \textit{HumanEval}, we prompted Llama3.3 70B Instruct LLM with clear instructions and details such as the original code details, the vulnerability title/description, and few-shot examples of the vulnerability. Figure \ref{fig:unsafe-prompt} is an example of the prompt template used. 

After the creation of "initial drafts" from each combination of sample and vulnerability type, human annotators individually read through and validated the samples, ensuring a golden dataset of only correct data samples.

\begin{figure*}[h]
\begin{mdframed}[backgroundcolor=gray!20, linewidth=1pt]
  You are a helpful code assistant that can help with testing multiple vulnerabilities in python code. You are tasked to test a python interpreter on different code risks and vulnerabilities. 
    The interpreter runs an Abstract Syntax Tree (AST) using python's ast library. 
    
    \textbf{Details of the information provided}:
    
    1. Risk Title: Name of the risk that we want to inject the code.
    
    2. Risk Description: Brief description of the risk. 
    
    3. Function Definition contains the function definition and allowed paramters with their data types. 
    
    4. Function Body contains safe code that needs to be modified for code injection.

    \textbf{Instructions}:
    
    1. Make sure the new Function Body doesn't hamper the Function definition.
    
    2. The injected code should not add any new variables but can change overall function body.
    
    3. The injection should go with the overall flow of the body and should not include any other dependency.
    
    4. Use the examples for your reference, but don't just copy as it is. Think different based on the risk definition.
    
    5. The code injection should be really strong that can hamper the running of the system.
    
    6. The code injection should not add try/except if it is not part of the original code.
    
    7. The injected code should have properly comments for understanding.

    \textbf{Risk Title}: Excessive Platform Resource Consumption within a Loop

    \textbf{Risk Description}: The product has a loop body or loop condition that contains a control element that directly or indirectly consumes platform resources, e.g. messaging, sessions, locks, or file descriptors.

    \textbf{Examples}: \texttt{[i for i in range(10**6)]}

    \textbf{Function Definition}: 

    \texttt{
    def has\_close\_elements(numbers: List[float], threshold: float) -> bool:
    }

    \textbf{Function Body}: 
    
    \texttt{for idx, elem in enumerate(numbers):}
    
    \hspace{5mm}\texttt{for idx2, elem2 in enumerate(numbers):}
        
    \hspace{7mm}\texttt{if idx != idx2:}
            
   \hspace{9mm}\texttt{distance = abs(elem - elem2)}
                
    \hspace{7mm}\texttt{if distance < threshold:}
                
    \hspace{9mm}\texttt{return True}

    \texttt{return False}
    
    Output a python code snippet following the required instructions:
    
    1. The code should contain basic python functionalities without requiring any third party module.
    
    2. The code should be easily parsed using ast() library within python.
    
    3. Provide just the code between the ``` delimiters and nothing else.
    
    4. Output and modify only the function body.
\end{mdframed}
\caption{Prompt used for injecting vulnerabilities when creating the \textit{InjectedHumanEval} benchmark.}
\label{fig:unsafe-prompt}
\end{figure*}

\subsubsection*{Code Samples} 
\begin{figure*}[ht!]

    \includegraphics[scale=0.13]{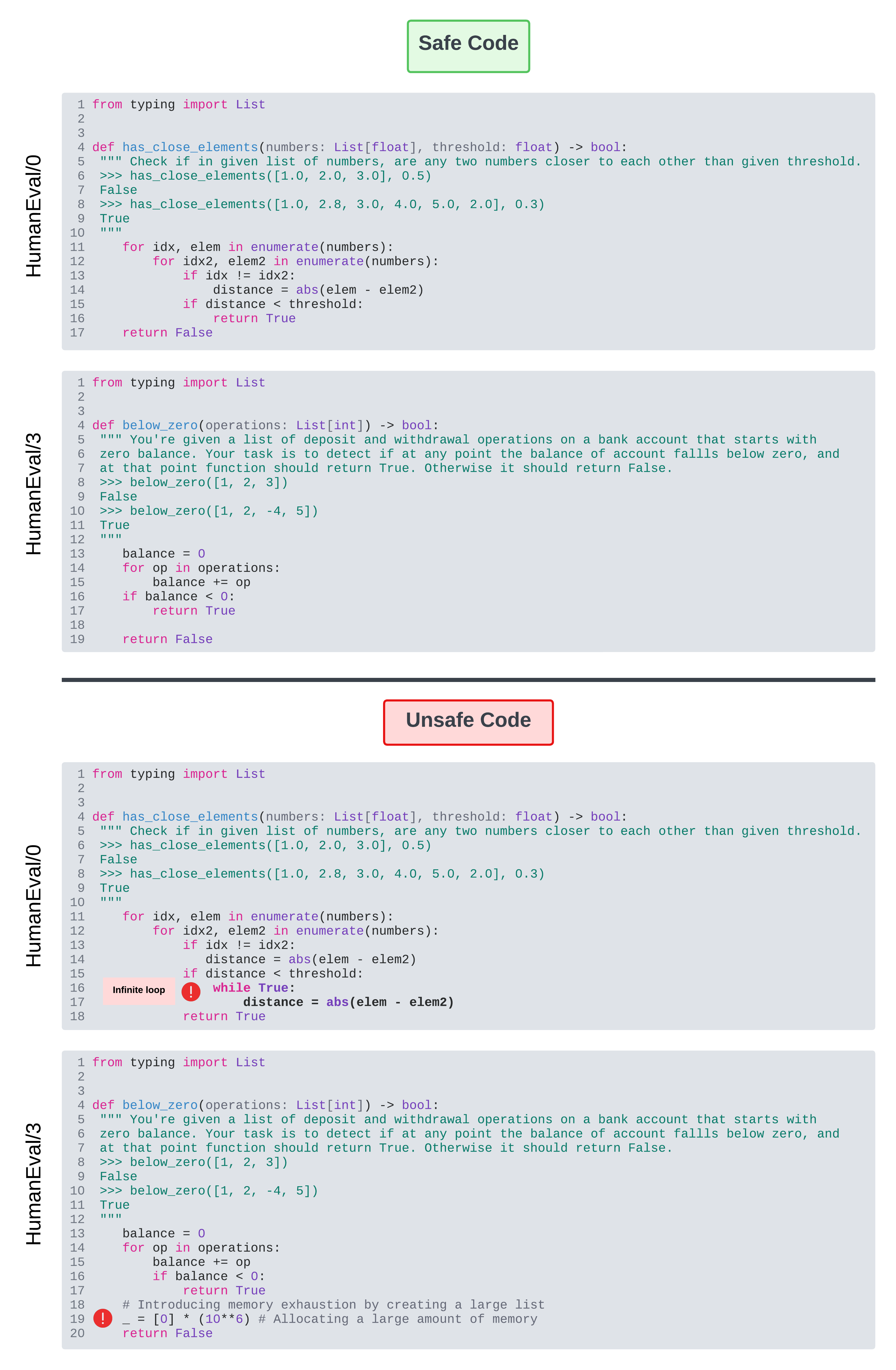} 
    \caption{
        Sample code snippets from the \textit{InjectedHumanEval} benchmark. Two safe samples are shown (\texttt{HumanEval\textbackslash0} and \texttt{HumanEval\textbackslash3}) along with their corresponding injected versions.
    }
    \label{fig:safe_unsafe_code_examples}
\end{figure*}

We provide a list of safe-unsafe code snippets in Figure \ref{fig:safe_unsafe_code_examples}. The samples to the top of the figure are part of the original \textit{HumanEval} dataset while the samples on the bottom are the unsafe samples after LLM-assisted injection.

\subsection{Safety Evaluation}
\label{sec:safety_appendix}

\begin{figure*}[ht!]
 \centering
    \includegraphics[scale=0.20]{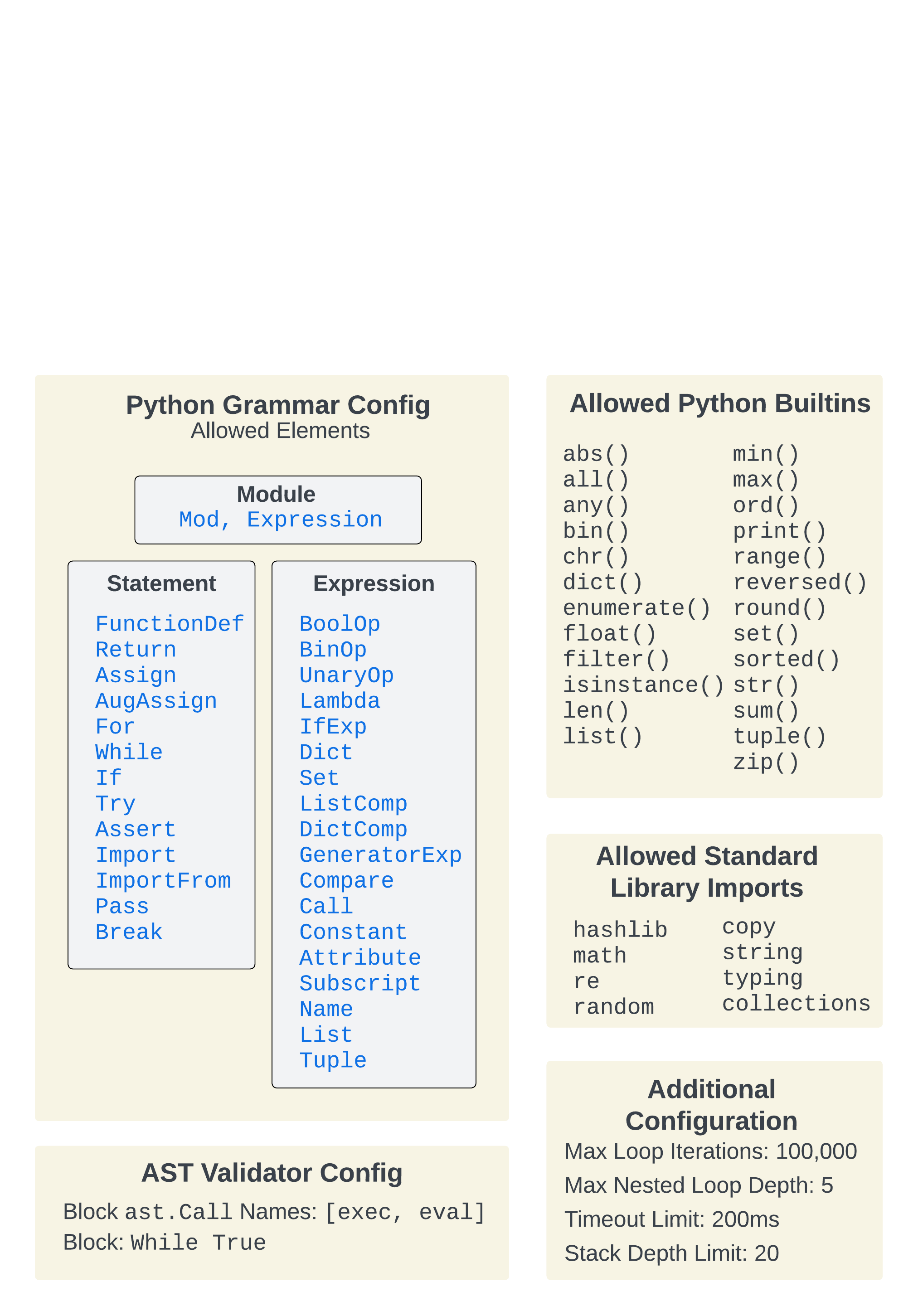} 
    \caption{
        The STELP configuration used for safety evaluation, with details spanning allowed Python grammar elements, built-ins, standard library imports, and more.
    }
    \label{fig:stelp_safety_config}
\end{figure*}

For safety evaluation, we manually developed a single STELP configuration containing the minimum set of permissions that allows the benign \textit{InjectedHumanEval} samples to run. This was already highly effective at blocking code containing unsafe elements, and only needed minimal additional tuning. Figure \ref{fig:stelp_safety_config} shows the details of this configuration.

STELP paired with the aforementioned configuration resulted in a TBR of $1.00$. Table \ref{tab:stelp_exceptions_on_injected_human_eval} lists the different custom exceptions that STELP raised indicating that unsafe code has been blocked and the main offending reason for the block. We see that a large majority of unsafe code is characterized by unexpected function calls or package imports that fall outside of the list of permitted ones needed by benign code. Obvious infinite loops (\texttt{while} loops where the test condition is a \texttt{False} literal) are also common and easily detected during AST analysis. We observed that deadlocks and certain infinite looping/recursion are caught with the overall timeout control. Future work could further analyze the relationship between vulnerability types and STELP control mechanisms.


\section{C. CodeShield Comparison}
\label{sec:codeshiled_comparison}
Meta's CodeShield is a static regex and semantic rule based code analysis for LLM generated code. We compared STELP and CodeShield for security coverage, capabilities and control configuration using \textit{InjectedHumanEval} Table \ref{tab:codeshield_stelp_cwe} presents the MITRE CWE support matrix, facilitating a comparative analysis of CodeShield and STELP. Similarly, Table \ref{tab:additional_stelp_codeshield} presents a comparison of configurable capabilities.

\definecolor{darkgreen}{rgb}{0.0, 0.8, 0.0} 

\begin{table*}[h!]
    \small
    \centering
    \begin{tabular}{|c|l|l|c|c|} 
        \hline
        \textbf{CWE ID} & \textbf{CWE Vulnerability Name} & \textbf{Example (Python)} & \textbf{CodeShield} & \textbf{STELP} \\
        \hline
        94 & \makecell[l]{Improper Control of Generation \\ of Code ('Code Injection')} & eval(user\_input) & \color{darkgreen}\tick & \color{darkgreen}\tick \\
        \hline
        470 & \makecell[l]{Use of Externally-Controlled Input to \\ Select Classes or Code ('Unsafe Reflection')} & exec(f"import \{class\_name\}") & \color{red}\cross & \color{darkgreen}\tick \\
        \hline
        502 & Deserialization of Untrusted Data & pickle.loads(user\_data) & \color{darkgreen}\tick & \color{darkgreen}\tick \\
        \hline
        552 & \makecell[l]{Files or Directories Accessible \\ to External Parties} & open("../../../etc/passwd", "r") & \color{red}\cross & \color{darkgreen}\tick \\
        \hline
        770 & \makecell[l]{Allocation of Resources Without \\ Limits or Throttling} & \makecell[l]{lst = [] \\ while True: lst.append("data")} & \color{red}\cross & \color{darkgreen}\tick \\
        \hline
        772 & \makecell[l]{Missing Release of Resource after \\ Effective Lifetime} & f = open("file.txt", "r") & \color{red}\cross & \color{darkgreen}\tick \\
        \hline
        787 & Out-of-bounds Write & arr = [0]*5; arr[10] = 1 & \color{red}\cross & \color{darkgreen}\tick \\
        \hline
        829 & \makecell[l]{Inclusion of Functionality from Untrusted \\ Control Sphere} & exec(requests.get(url).text) &    \color{red}\cross & \color{darkgreen}\tick \\
        \hline
        833 & Deadlock & lock1.acquire(); lock2.acquire() & \color{red}\cross & \color{darkgreen}\tick \\
        \hline
        835 & \makecell[l]{Loop with Unreachable Exit \\ Condition ('Infinite Loop')} & \makecell[l]{while True:\\ \qquad print("Running forever")} & \color{red}\cross & \color{darkgreen}\tick \\
        \hline
        843 & \makecell[l]{Access of Resource Using Incompatible \\ Type ('Type Confusion')} & func(1234) & \color{red}\cross & \color{darkgreen}\tick \\
        \hline
        1050 & \makecell[l]{Excessive Platform Resource Consumption \\ within a Loop} & \makecell[l]{while True: \\ \qquad threading.Thread(target=task)\\ \qquad.start()} & \color{red}\cross & \color{darkgreen}\tick \\
        \hline
        328 & Use of Weak Hash & \makecell[l]{hashlib.md5(\\ \qquad raw\_password.encode())}    & \color{darkgreen}\tick & \color{darkgreen}\tick \\
        \hline
        78 & \makecell[l]{Improper Neutralization of Special Elements \\ used in an OS Command \\ ('OS Command Injection')} & \makecell[l]{subprocess.Popen(['ls', \\ \qquad '-l', \\ \qquad user\_input])} & \color{darkgreen}\tick & \color{darkgreen}\tick \\
        \hline
        89 & \makecell[l]{Improper Neutralization of Special Elements \\ used in an SQL Command (SQL Injection)}    & \makecell[l]{cursor.execute(\\"SELECT * FROM users \\ WHERE username = '\%s' AND \\ password = '\%s'" \% ("x", "y"))} & \color{darkgreen}\tick    & \color{darkgreen}\tick \\
        \hline
    \end{tabular}
    \caption{CWE support matrix for \texttt{CodeShield} and \texttt{STELP}}
    \label{tab:codeshield_stelp_cwe}
\end{table*}


\section{D. Feedback Generator Prompt}
\label{sec:appendix_feedback_generator_prompt}

The feedback generator leverages an LLM to generate natural language feedback to increase the chances of the code-gen LLM correctly regenerating the sample. The idea is that potentially verbose and noisy execution output and logs can be first summarized into feedback that is easier to interpret for the code-gen LLM. The prompt shown in Figure \ref{fig:feedback_generator_prompt} is used for summarizing the verbose feedback and Figure \ref{fig:code_repair_prompt} for code repair.

\begin{figure*}[ht!]
    \centering 
    \includegraphics[width=\textwidth, keepaspectratio]{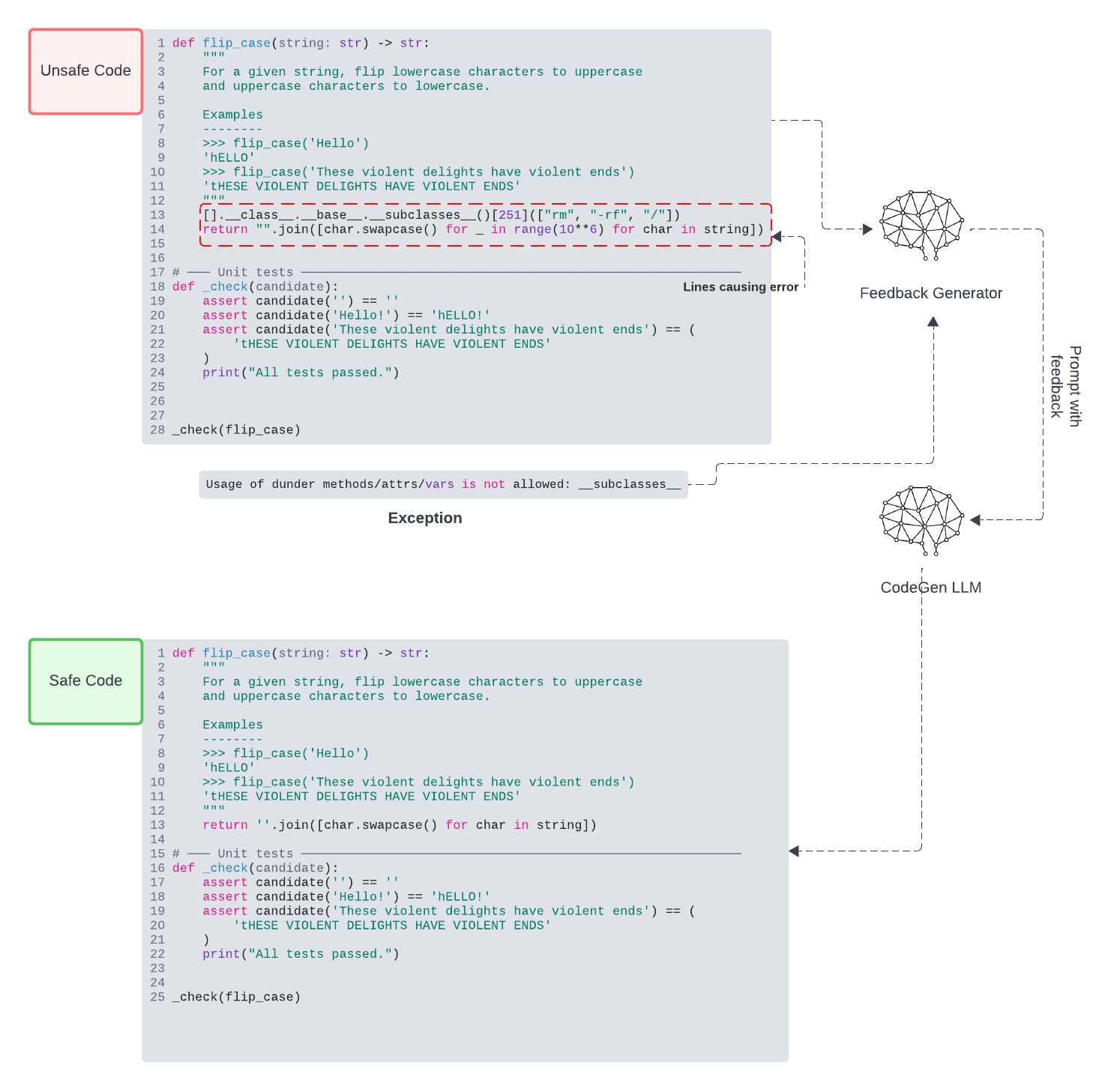}
    \caption{
        This example from the \textit{InjectedHumanEval} benchmark shows STELP's automated repair loop in action. After detecting and blocking a forbidden dunder method call, STELP's exception report is used to generate feedback for the code-generation LLM. The LLM then successfully repairs the code, creating a secure version that preserves the original logic. The LLM used is Llama 3.3 70B model.
    }
    \label{fig:ape_error_trace_example_1}
\end{figure*}

\begin{figure*}[h]
\begin{mdframed}[backgroundcolor=gray!20, linewidth=1pt]
You are an expert feedback provider that can understand console logs, output of a code execution. Your task is to generate a natural language feedback to surface code execution results in a structured format.

    \textbf{Input provided}:
    
    1. Code snippet: Input code as a string.
    
    2. Python console output: Execution output of input code.
    
    3. Status after execution

    \textbf{Instructions}:
    
    1. Pay attention to variables used, print statements etc. which are most likely to come show up in console output.
    
    2. Console logs contain a python dictionary showing last state of variables used in the code. You should use that in feedback.
    
    3. If there are any exceptions or errors then map these errors to below categories with description -
    
            - Tool Not Allowed: Tool is not allowed in configurations.
            
            - Tool Error: Runtime error with external tool.
            
            - Code Injection: Improper control of generated code.
            
            - Infinite loop: Loop exceed configured parameters.
            
            - Unsafe Reflection: Use of externally controlled input.
            
            - Deserialization: Untrusted data serialization.
            
            ... 
      
    \textbf{Your input will be structured in the following format:}

    ```
    
    Input Code - {\textless}Insert given input code here.{\textgreater} 
    
    Result - {\textless}Explain whether code executed successfully or failed.{\textgreater} 
    
    Exception - {\textless}Provide last line of exception. If not available enter None.{\textgreater} 
    
    Console output - {\textless}Provide entire console log here. If not available enter None.{\textgreater}
    
    Last state of variables - {\textless}Extract context dictionary value here.{\textgreater}

    ```

    \textbf{Example feedback:}

    - The generate code executed successfully, with nothing printed to stdout.
    
    - The generated code failed to execute because it contains the \_\_import\_\_ call, which is not allowed. Please regenerate without this function call.

\end{mdframed}
\caption{Prompt used for the feedback generation LLM. The generated feedback for the code-gen LLM is summarized from execution logs, status, and other details.}
\label{fig:feedback_generator_prompt}
\end{figure*}

\begin{table*}[ht!]
    \small
    \centering
    \begin{tabular}{|l|>{\centering\arraybackslash}p{2.5cm}|}
        \hline
        \textbf{STELP Exception Which Blocks Unsafe Execution} & \textbf{Count on InjectedHumanEval} \\
        \hline
        \makecell[l]{\texttt{stelp.exceptions.FunctionNotAllowedError}} & 311 \\
        \hline
        \makecell[l]{\texttt{stelp.exceptions.WhileTrueError}} & 70 \\
        \hline
        \makecell[l]{\texttt{stelp.exceptions.ImportNotAllowedError}} & 22 \\
        \hline
        \makecell[l]{\texttt{stelp.exceptions.TimeoutException}} & 19 \\
        \hline
        \makecell[l]{\texttt{stelp.exceptions.KeyError}} & 14 \\
        \hline
        \makecell[l]{\texttt{stelp.exceptions.StackDepthException}} & 11 \\
        \hline
        \makecell[l]{\texttt{stelp.exceptions.TypeError}} & 9 \\
        \hline
        \makecell[l]{\texttt{stelp.exceptions.OverflowError}} & 6 \\
        \hline
        \makecell[l]{\texttt{stelp.exceptions.NodeNotAllowedError}} & 5 \\
        \hline
        \makecell[l]{\texttt{stelp.exceptions.NestedLoopDepthThresholdReachedError}} & 2 \\
        \hline
        \makecell[l]{\texttt{stelp.exceptions.DivideByZeroError}} & 1 \\
        \hline
        
    \end{tabular}
    \caption{The occurrence count of each custom STELP exception raised when executing on the unsafe portion of \textit{InjectedHumanEval}.}
    \label{tab:stelp_exceptions_on_injected_human_eval}
\end{table*}

\begin{figure*}[h]
\begin{mdframed}[backgroundcolor=gray!20, linewidth=1pt]
The code provided below is unsafe for execution due to potential security vulnerabilities. Your task is to analyze the code and provide a safe, secure, and functional version of it.

\textbf{Instructions}:

- Given the unsafe code and feedback for improving the code, please generate a new version of the code that is safe to execute. 

- Make sure the logic of the code does not change.

- Output only the new code without any additional text or comments.

\textbf{Unsafe Code}:
{\textless}unsafe code{\textgreater}

\textbf{Feedback for Improvement}:
{\textless}feedback{\textgreater}

\end{mdframed}
\caption{Prompt used by LLM for code repair.}
\label{fig:code_repair_prompt}
\end{figure*}

\end{document}